\documentclass[runningheads]{llncs}
\usepackage{graphicx}
\usepackage{multirow}
\usepackage{amsmath,amssymb,amsfonts}
\usepackage{url}
\usepackage{listings}

\newcommand{\sig}[1]{{\small\textsf{{#1}}}}

\usepackage{array,booktabs}

\newcolumntype{P}[1]{>{\centering\arraybackslash}p{#1}}

\newtheorem{defi}{Definition}

\newcommand{\Comment}[1]{}

\begin{document}
\title{Logically Sound Arguments for the Effectiveness of ML Safety Measures}
%\title{Towards Logical Argumentation for Mediating Object Detection Safety Concerns}
%\title{Logical Argumentation for Mediating Object Detection Safety Concerns via Conservative Post-Processing}
%\subtitle{(ETAPS'22 - FASE NIER track)}

\author{Chih-Hong Cheng\and
Tobias Schuster\and
Simon Burton}
%

\institute{
Fraunhofer IKS, Munich, Germany\\
\email{\{firstname.lastname\}@iks.fraunhofer.de}}

%\authorrunning{C.-H. Cheng et al.}
%\titlerunning{Logically Sound Arguments for the Effectiveness of ML Safety Measures}

%
\maketitle              % typeset the header of the contribution
%

%\vspace{-2mm}

\vspace{-5mm}

\begin{abstract}

We investigate the issues of achieving sufficient rigor in the arguments for the safety of machine learning functions. By considering the known weaknesses of DNN-based 2D bounding box detection algorithms, we sharpen the metric of imprecise pedestrian localization by associating it with the safety goal. The sharpening leads to introducing a conservative post-processor after the standard non-max-suppression as a counter-measure. We then propose a semi-formal assurance case for arguing the effectiveness of the post-processor, which is further translated into formal proof obligations for demonstrating the soundness of the arguments. Applying theorem proving not only discovers the need to introduce missing claims and mathematical concepts but also reveals the limitation of Dempster-Shafer's rules used in semi-formal argumentation.

%\keywords{safety argumentation  \and SOTIF \and autonomous driving}

\end{abstract}

\vspace{-10mm}

\section{Introduction}

Ensuring safety has been considered one of the critical barriers in realizing autonomous driving functionalities. While newly developed safety standards such as ISO~21448 SOTIF~\cite{SOTIF} address hazards directly caused by the intended functionality such as performance limitations of perception modules, it only provides a high-level methodology for safety assurance, and how one implements the process and the resulting evidence construction are left for interpretation.

In this paper, we study how to achieve sufficient rigor in the safety argumentation for \emph{Deep Neural Networks} (DNN)~\cite{lecun2015deep} using an example of 2D object detection. In contrast to classical software, deep neural networks offer a more direct approach to quantify the risk associated with the software. Nevertheless, the first barrier towards  scientific rigor occurs due to standard quality attributes for evaluating the performance of DNNs having very obscure connections to safety. We demonstrate with a simple example that the widely used Intersection-over-Union (IoU) metric may not be positively correlated to simple safety goals such as avoiding collisions. We then use a simple criterion to evaluate the DNN perception module against their impact on safety. For issues caused by localization imprecision, the simple and direct criterion leads to an introduction of a safety post-processor whose configuration is determined after training is completed.

Subsequently, we utilize the semi-formal Goal Structuring Notation (GSN)~\cite{kelly2004goal} to detail arguments why introducing such a post-processor is sufficient to achieve the safety goal. Due to the clarity of the newly introduced safety metrics, we can \emph{formulate all semi-formal specifications into logical formulae}, and try to formally deduce that the arguments are sufficient to derive the goal using interactive theorem proving~\cite{owre1996pvs}. This deduction step is critical to achieve the desired rigor for safety-critical ML systems - There exists a fundamental difference between claims (supported by a qualitative, semi-formal argument) and proven conjectures.  In our example, the deduction fails unless we introduce (1) additional problem-specific claims and (2) mathematical concepts as lemmas. This initial result demonstrates the possibility of bringing mathematical rigor in the safety argumentation of DNNs, provided that the definition of safety and the associated evaluation function are made available. 

Finally, moving towards quantitative argumentation, the failed logical deduction highlights the \emph{limitation of existing semi-formal approaches in evidence combination} based on Dempster-Shafer's evidence theory~\cite{sentz2002combination,idmessaoud2020belief}.  Precisely, when using GSNs in the SOTIF context, the computed probability based on evidence combination shall be viewed as a claim or as an upper bound. The corollary is that quantitative safety assessment of machine learning functions is only valid when the connection between low-level arguments and high-level goals is sound through logical deduction. 

The rest of the paper is structured as follows. After summarizing related work in Section~\ref{sec.related.work}, in Section~\ref{sec.sotif} we provide the example under analysis and present the method of addressing functional insufficiencies. Subsequently, in Section~\ref{sec.logical.argumentation} we describe how a logical argumentation can be established as well as issues in moving towards a quantitative argumentation. Finally, we conclude in Section~\ref{sec.conclusion} by outlining further research opportunities.

%\vspace{-4mm}
\section{Related Work}\label{sec.related.work}
%\vspace{-4mm}

The safety of deep neural networks in the context of autonomous driving is currently under active research. On the methodology side, many results on safety argumentation use semi-formal/structural notations with variations on argumentation strategy or used metrics (to list a few~\cite{burton2017making,zhao2020safety,jia2021framework,salay2021missing}). These results have their merits but overall fail to provide a sound logical connection between low-level claims and high-level goals. The exercise demonstrated in this paper clearly shows the benefits and challenges of precisely formulating all claims and goals into logical formulae. Many existing results~\cite{cyra2011support,yuan2017subjective,wang2018confidence,idmessaoud2020belief} consider connecting semi-formal argumentation with Dempster-Shafer theory and apply evidence combination using rules such as Dempster's~\cite{dempster1968upper} or Yager's~\cite{yager1987dempster} rules. However, as pointed out in this paper, the translation and the generated evidence shall only be viewed as claims due to a lack of logical derivation.

Within the research community, numerous testing coverage criteria for deep neural networks have been proposed (see~\cite{huang2020survey} for a survey of recent work). While these proposals are scientifically novel,  it is nevertheless important to integrate these test coverage criteria or quality attributes into the overall safety argumentation. Many test criteria (e.g., neuron coverage~\cite{pei2017deepxplore}) proposed by the software engineering community admittedly lack the concrete connection to commonly seen dependability attributes. On the other side, metrics used in the machine learning community such as IoU can lead to counter-intuitive results (e.g., higher IoU does not imply being safer). The lack of connection between existing metrics and safety can easily be detected and fixed when  formalizing the high-level goals using logic, followed by a formal deduction exercise.

Finally, for the post-processing algorithm developed in this paper, the closest idea similar to ours is a recent work by Cheng~\cite{cheng2020safety} that replaces the standard non-max-suppression post-processing algorithm with a new one using non-max-inclusion. Although in~\cite{cheng2020safety} the new post-processing algorithm is connected with the safety goal of avoiding false negatives, there is no (logical) derivation how the newly developed post-processing contributes to the safety goal - a common weakness that appears in many existing  research results. 

\vspace{-2mm}

\section{Addressing Insufficiencies in Object Detection}~\label{sec.sotif}

\vspace{-5mm}

In this section, we consider an example of vision-based 2D object detection using deep neural networks. To ease understanding, some of the statements or assumptions are simplified to establish the formal proof.  Readers may extrapolate the ideas stated in this section and integrate various corner cases that one should consider when encountering real-world scenarios.

\vspace{-3mm}
\paragraph{(Background)} Given an image, a 2D object detector generates \emph{bounding boxes} around the area where it considers an existing object (e.g., pedestrian) with high probability. To train such an object detector, one provides a \emph{data set}; each data point in the data set is an image together with its associated ground truth bounding boxes. Ideally, the data set should be a \emph{good approximation} of the \emph{operational design domain} (ODD) where the object detector is used. %The  data set may further be partitioned into multiple sets to perform training, hyper-parameter tuning and understanding the generalization error. 

\vspace{-3mm}
\paragraph{(Understanding the meaning of safe perception)} For autonomous vehicles, one  hazardous scenario occurs when an un-occluded pedestrian within a certain distance is not properly identified. It is necessary  to precise the definition of ``properly identifying a pedestrian", as in the SOTIF process, one needs to make sure that the probability of improper identification is sufficiently low. Very commonly, one then proceeds by associating ``proper detection" with quality attributes / performance metrics widely used in machine learning.

\begin{figure}[t]
\includegraphics[width=0.9\textwidth]{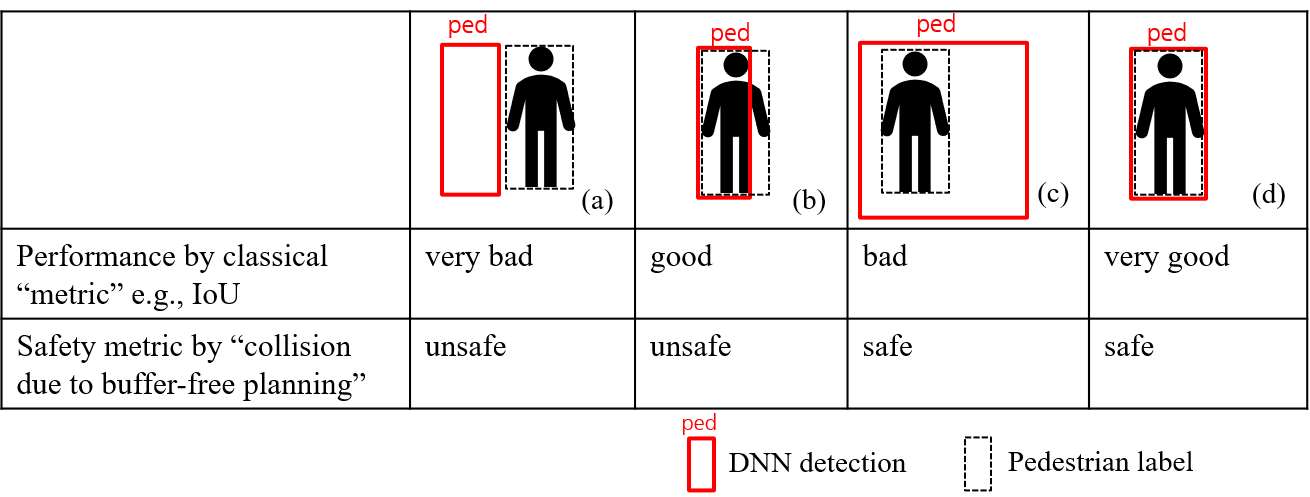}
\centering
\vspace{-4mm}
\caption{Standard performance metrics such as IoU may be incorrectly related to safety}
\label{fig.metrics}
\end{figure}

\vspace{1mm}
Nevertheless, as this association step is commonly done informally, the ambiguity can lead to severe  problems in safety. For example, consider the four sub-cases listed in Figure~\ref{fig.metrics}, where we show the DNN prediction and the ground truth label, both as bounding boxes. Between a prediction bounding box and a label bounding box, the commonly used \emph{Intersection-over-Union (IoU)} metric computes the area of overlap (of two bounding boxes) against the area over the union (of two bounding boxes). One can derive that the computed IoU values for Figure~\ref{fig.metrics}(a) and Figure~\ref{fig.metrics}(d) equal~$0$ and~$1$, respectively. For Figure~\ref{fig.metrics}(b), the computed IoU is larger than~$0.5$; for   Figure~\ref{fig.metrics}(c), the computed IoU is smaller than~$0.5$. Subsequently, when one uses IoU as an evaluation metric, one needs to \emph{argue how the computed IoU value is related to safety}. Consider a  setup where there is no safety buffer associated with the motion planning algorithm, then the situation in Figure~\ref{fig.metrics}(b) is not safe, as the vehicle may assume that the area outside the prediction corresponds to drivable free-space, thereby inducing the risk of collision. This contrasts with Figure~\ref{fig.metrics}(c), where although the DNN-predicted bounding box is large, a buffer-free motion planner that never encroaches the prediction bounding box can never hit the pedestrian. Therefore, under specific assumptions on the capability of the motion planner, the IoU has no direct correlation with safety. 
In this example, one possible definition of safe detection   can be informally\footnote{We keep the statement in Definition~\ref{def:safe.detection} informal for simplicity purposes; readers are encouraged to sharpen it by mathematically defining the meaning of bounding box and the concept of area containment and additional conditions such as partial occlusion or size information (reflecting the distance measure).} stated using Definition~\ref{def:safe.detection}.

\vspace{-1mm}
\begin{defi}[Safe DNN pedestrian detection]~\label{def:safe.detection} A DNN pedestrian detection is safe, if the DNN-predicted bounding box  strictly contains the labeled bounding box of the pedestrian.
\end{defi}

\vspace{-5mm}
\paragraph{(Addressing imprecise detection as functional insufficiency)} Based on Definition~\ref{def:safe.detection}, one can take an object detector and understand if the generated prediction is safe. Unsafe predictions may  further be categorized by their demonstrated behaviors.  For each category, one should design a corresponding counter-measure to inhibit the likelihood of occurrence. Here we focus on addressing the behavior of \emph{imprecise localization}, i.e., a pedestrian is detected by the object recognition module, but the imprecision of the detector can not fully cover the bounding box, similar to the examples in Figure~\ref{fig.metrics}(a) and~(b). The consequence of using standard performance metrics such as mean-average precision (mAP) is that one needs to \emph{perfect the performance to achieve safety}. Unfortunately, the state-of-the-art DNN-based object detectors as of 2021~\cite{dai2021dynamic} setting a benchmark for the COCO object detection dataset~\cite{lin2014microsoft} only reaches an mAP of~$78.5\%$, making it still far away from claiming the performance to be ``perfect". However, we demonstrate in the below paragraph that we have a simpler way of improving the overall perception module. 

\begin{figure}[t]
\centering
\includegraphics[width=0.6\textwidth]{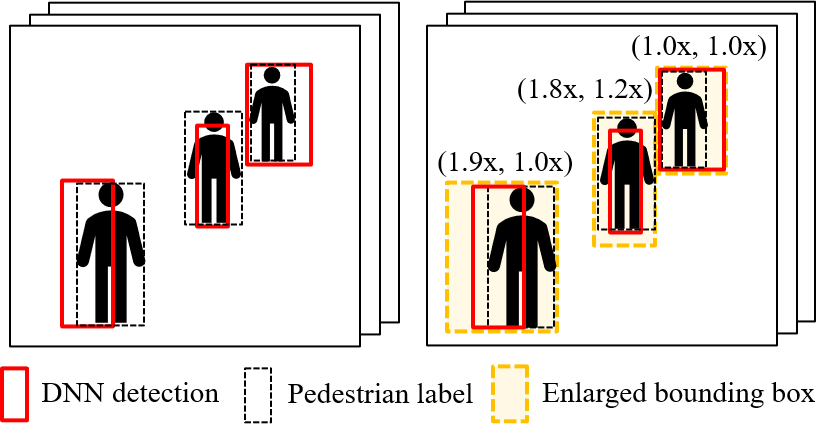}
\vspace{-2mm}
\caption{Enlarging the bounding box to enable full containment}
\label{fig:enlargement}
\vspace{-2mm}

\end{figure}

\begin{figure}[t]
\centering
\includegraphics[width=0.9\textwidth]{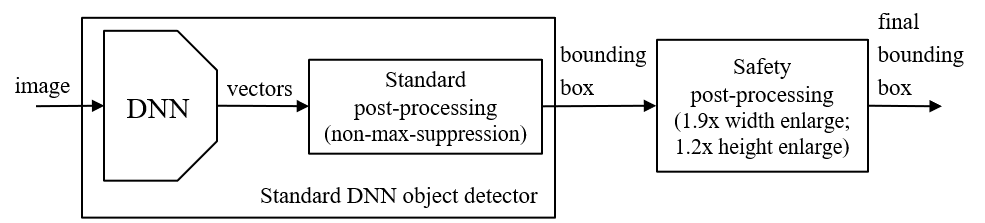}
\vspace{-2mm}
\caption{Function modification by adding safe post-processing modules}
\label{fig:safe.post.processing}
\end{figure}

\vspace{-3mm}
\paragraph{(Conservative Safety-aware Post-Processing)} Given that a DNN-based perception module is unable to generate bounding boxes that fully cover the pedestrian labels, instead of retraining the DNN module, we perform \emph{function modification} as detailed below. The intuition is that by observing and measuring the degree of localization imprecision from the training data, one can conservatively compensate it in operation by enlarging the bounding box. 

%\vspace{-3mm}

\begin{enumerate}
\item For each image collected in the training data set, and for each predicted bounding box that does not fully cover the pedestrian label, measure the minimum enlargement ratio required to enclose the bounding box label. Consider the example in Figure~\ref{fig:enlargement}. For the bounding box prediction in the bottom left, to fully cover the label, one should enlarge the width of the bounding box by~$1.9\times$. For the prediction in the center of the image, to fully cover the label, one should enlarge the width of the bounding box by~$1.8\times$ and enlarge the height by~$1.2\times$. 
    
\item Learn the required ratio from the previous step, in order to be larger than all previously computed ratios while considering variations in the training data distribution. Consider again the example in Figure~\ref{fig:enlargement}, a simple ``mechanical learning" can be to memorize the maximum ratio for all images in the training data set and for all bounding boxes analyzed in the previous step. In Figure~\ref{fig:enlargement},
the maximum width expansion ratio equals~$\sig{max}\{1.9, 1.8, 1.0\} = 1.9$, and  the maximum height expansion ratio equals $\sig{max}\{1.0, 1.2, 1.0\} = 1.2$. 
    
\item Finally, add another post-processing unit after the standard post-processing unit, as illustrated in Figure~\ref{fig:safe.post.processing}. During runtime, whenever a bounding box is generated, the post-processor always enlarges the predicted bounding box by the ratio stored in the previous step. For the example in Figure~\ref{fig:enlargement}, for every bounding box predicted in run time, the additional post-processor  enlarges  its width by $1.9\times$ and its height by $1.2\times$ to create the final prediction. 
\end{enumerate}

\begin{figure}[t]
\includegraphics[width=\textwidth]{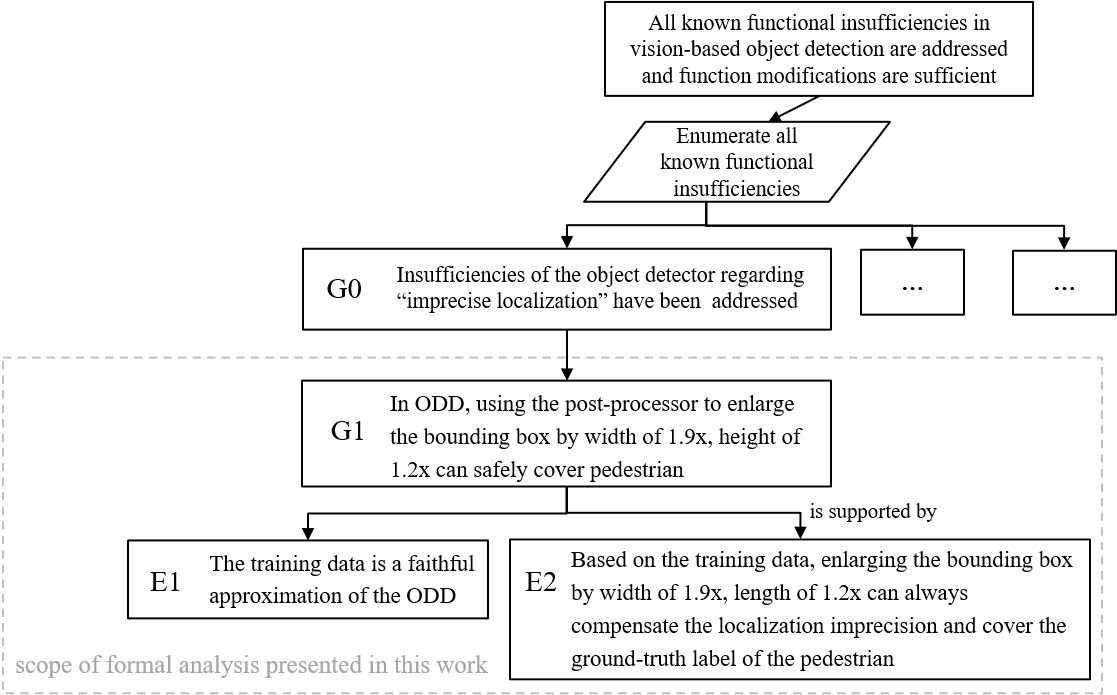}
\centering
\vspace{-2mm}
\caption{Structuring the argument for safety, as a first attempt}
\label{fig.problematic.derivation}
\end{figure}

%\vspace{-5mm}

\section{From Structural to Logical Argumentation}~\label{sec.logical.argumentation}

%\vspace{-5mm}

For the approach presented in the previous section applied on the example in Figure~\ref{fig:enlargement}, Figure~\ref{fig.problematic.derivation} shows the first attempt to depict the structural argumentation connecting the goal~G1 with two supporting arguments~E1 and~E2. The underlying idea is that the prediction can always cover the pedestrian (G1), provided that

\vspace{-2mm}
\begin{itemize}
    \item  enlarging bounding boxes  demonstrates its sufficiency (E2) in-sample, and
    \item the training data is a good approximation of the ODD (E1), thereby creating a formal connection between in-sample and out-of-sample performance. 
\end{itemize}
\vspace{-2mm}

Subsequently, we perform a manual translation from informal specification to logical specification. Table~\ref{table:formulation} shows the translated formula, and we refer readers to the appendix for formal specifications described using PVS~\cite{owre1992pvs}. The meaning of each predicate and function should be self-explanatory; here we only briefly explain its usage without giving a formal definition: $\sig{Training}(d)$ returns \sig{true} iff an image~$d$ is within the training data set; $\sig{ODD}(d)$ returns \sig{true} iff an image~$d$ is in the ODD. Given an image~$d$, $\sig{DNN}(d)$, $\sig{label}(d)$ and $\sig{ground\_truth}(d)$ returns the predicted, labelled, and ground-truth bounding boxes of pedestrians. Function $\sig{Enlarge}_{(1.9,1.2)}$ is the post-processor that enlarges each of the bounding boxes generated by the DNN prediction. Finally, $\sig{Cover}(S1, S2)$ returns $\sig{true}$ iff bounding boxes in~$S1$ can cover bounding boxes in~$S2$.

For~E1, there can be multiple ways of interpreting ``faithful approximation" such as Probably Approximately Correct (PAC) learning with bounded generalization errors. Here for the ease of understanding, we use an extremely strong condition on \emph{behavior equivalence}, i.e., the behavior of the DNN demonstrated in the ODD will be the same as the behavior demonstrated in the training data set, applicable for every evaluation function $\sig{Behavior}$, under possible local transformations over the output of the DNN ($\sig{F}_1$) and the data point ($\sig{F}_2$). By translating~E1, E2, and G1 into logical formulae, one can perform a deduction and infer that \emph{even under such a strong generalizability condition}, E1 and~E2 alone are \emph{not sufficient} to derive~G1.

\vspace{-1mm}

\begin{itemize}
    \item Within the training data, it is also important that the label of a pedestrian actually surrounds the corresponding pedestrian. This is characterized by~E3 as stated in Table~\ref{table:formulation.additional}. 
    \item To allow mechanical deduction, one needs to additionally add a \emph{mathematical concept} (transitivity rule) for \emph{area containment}, stating that if~A covers~B, and if~B covers~C, then~A also covers~C. This is characterized by~E4 as stated in Table~\ref{table:formulation.additional}.
    
\end{itemize}

\begin{table}[t]
\begin{small}
\centering
\bgroup
\def\arraystretch{1.2}
\begin{tabular}{|P{9mm}|P{113mm}|}
\hline
Index & Informal description and the corresponding formal specification\\
\hline
\multirow{2}{*}{E1} & \; The training data is a faithful approximation of the ODD  \\
\cline{2-2}
&  $\forall \sig{Behavior}, \sig{F}_1, \sig{F}_2: (\forall d :
\sig{Training}(d) \rightarrow \sig{Behavior}(\sig{F}_{1}(\sig{DNN}(d)), \sig{F}_2(d))) $ \\
&
$ \qquad\qquad\qquad\qquad \rightarrow (\forall d_{op} :
\sig{ODD}(d_{op}) \rightarrow \sig{Behavior}(\sig{F}_{1}(\sig{DNN}(d_{op})), \sig{F}_2(d_{op})))$ \\
\hline
\multirow{2}{*}{E2} & Based on the training data, enlarging the bounding box by width of $1.9\times$, height of $1.2\times$ can always cover the ground-truth label of the pedestrian \\
\cline{2-2}
& $\forall d : \sig{Training}(d) \rightarrow \sig{Cover}(\sig{Enlarge}_{(1.9,1.2)}(\sig{DNN}(d)), \sig{label}(d))$ \\
\hline
\multirow{2}{*}{G1} &  In ODD, enlarging the bounding box by width of $1.9\times$, height of $1.2\times$ can safely cover the pedestrian \\
\cline{2-2}
& $\forall d_{op} : \sig{ODD}(d_{op}) \rightarrow \sig{Cover}(\sig{Enlarge}_{(1.9,1.2)}(\sig{DNN}(d_{op})), \sig{ground\_truth}(d_{op}))$ \\
\hline
\end{tabular}
\vspace{1mm}
\caption{Translating informal arguments in Figure~\ref{fig.problematic.derivation} into logical formula}
\egroup
\end{small}
\vspace{-3mm}
\label{table:formulation}
\end{table}

\begin{table}[t]
\begin{small}
\centering
\bgroup
\def\arraystretch{1.2}
\begin{tabular}{|P{9mm}|P{113mm}|}
%\hline
%Index & Informal description and the corresponding formal specification\\
\hline
\multirow{2}{*}{E3} & Based on the training data, ground-truth label of the pedestrian always strictly contains the pedestrian \\
\cline{2-2}
& $\forall d : \sig{Training}(d) \rightarrow \sig{Cover}( \sig{label}(d), \sig{ground\_truth}(d))$ \\
\hline
\multirow{2}{*}{E4} & If A covers B, and B covers C, then A covers C \\
\cline{2-2}
& $\forall A, B, C : (\sig{Cover}(A, B) \wedge \sig{Cover}(B, C)) \rightarrow \sig{Cover}(A, C) $  \\
\hline
\end{tabular}
\egroup
\end{small}
\vspace{2mm}
\caption{Additional arguments}
\label{table:formulation.additional}
\vspace{-3mm}
\end{table}

\vspace{-2mm}
\noindent The formally stated specification in PVS~\cite{owre1992pvs} is checked mechanically using interactive theorem proving to derive that~G1 can be deduced by  arguments \{E1,  \ldots, E4\}. Intuitively, 
the logical deduction of proving~G1 proceeds by first creating a new lemma~E5, stating that within the training data set, as every enlarged bounding box covers the label (E2), and as every label covers the ground truth location of the pedestrian (E3), the enlarged bounding box can (due to transitivity E4) cover the ground truth location of the pedestrian.

\vspace{-3mm}

\begin{equation*}
\forall d : \sig{Training}(d) \rightarrow \sig{Cover}(\sig{Enlarge}_{(1.9,1.2)}(\sig{DNN}(d)), \sig{ground\_truth}(d))\tag{E5}
\end{equation*} 

Subsequently, use E1 by instantiating $\sig{Behavior}$, $\sig{F}_1$, and $\sig{F}_2$ with $\sig{Cover}$, $\sig{Enlarge}_{(1.9,1.2)}$, and $\sig{ground\_truth}$ respectively. Then together with~E5, one derives~G1. 

\vspace{-2mm}

\paragraph{(Towards quantitative argumentation)} The example demonstrated in Figure~\ref{fig.problematic.derivation} also provides an example on the limitation of using semi-formal notations for quantitative argumentation suggested by earlier results. Even when E1 and E2 have $100\%$ confidence of being correct, one may be tempted to use evidence combination theory such as Dempster's rule on structural argumentation~\cite{cyra2011support,yuan2017subjective,wang2018confidence,idmessaoud2020belief} and to conclude that~G1 holds with~$100\%$ confidence. Nevertheless, based on the logical deduction, G1 does not hold without E3. This implies that the result of evidence combination can only be counted as an upper bound.  

%\vspace{-2mm}

\section{Concluding Remarks}~\label{sec.conclusion}

%\vspace{-4mm}

The motivation of this work is based on our belief that engineering safety-critical AI systems should be based on a scientifically grounded approach that (1) establishes a well-thought process, (2) uncovers all hidden assumptions in the design, (3) establishes evidence via clearly comprehensible KPIs connected to safety goals and finally, (4) designs mechanisms to allow continuous product improvement. As a key takeaway, we realized that some of the currently executed ``best practices" in engineering DNN-based perception systems should not be ``taken for granted", as they can surely be sharpened by investing mild efforts in translating all semi-formal claims and concepts into logical formulae. 

Throughout the study, we were able to design new safety-aware performance metrics that can be directly connected to safety goals, suggest a function modification (safety-aware post-processor) as a direct counter-measure over the safety metric, and finally, uncover hidden assumptions that are missing in the semi-formal argumentation. We also noted the risk of moving from qualitative argumentation towards quantitative argumentation without a sound logical foundation. Although performing such a translation and documenting them require additional efforts, in a continuous engineering paradigm being predominant in the current autonomous driving development, one maintains the argumentation tree and performs only partial argument updates. The cost can be even-out when counting multiple sprints being executed until deployment, while the benefits of introducing logic and precision in reasoning/deduction may be substantial in improving system safety.  

This study is an initial step towards rigorous-yet-efficient engineering of trustworthy learning-enabled systems. The following directions are currently under development: (i) Expand the example by considering other types of performance limitations. (ii) Investigate synthesis methods to automatically deduce the weakest condition  to enable automatic theorem proving. (iii) Automate the process of evidence maintenance via tool support~\cite{beyene2018evidential,carlan2020fasten}. (iv) Perform a detailed study on evidence combination specialized for deep neural networks, by considering correlation and weight-difference between pieces of evidence.  (v) Relax and refine the training-to-ODD generalization argument with measurable sub-evidences. (vi) Use automatic code generation to generate provably correct runtime monitors.

\subsubsection*{Acknowledgement} This work is funded by the Bavarian Ministry for Economic Affairs, Regional Development and Energy as part of a project to support the thematic development of the Fraunhofer Institute for Cognitive Systems.

\bibliographystyle{abbrv}
%\bibliography{ref}

\newpage

\section*{Appendix}

\subsection*{Part A. Specification in PVS}

\begin{lstlisting}
FASE22_Perception_full  : THEORY

  BEGIN

  BB : Type  % Bounding box
  IMG : Type % Image
  BEHAVIOR: TYPE = [BB, BB -> bool]
  F1: TYPE = [BB -> BB]
  F2: TYPE = [IMG -> BB]

  cover:  [BB, BB -> bool]

  in_odd: [IMG -> bool ]
  in_training_set: [IMG -> bool ]

  dnn: [IMG -> BB]
  enlarge_by_post :  [BB -> BB]
  label:  [IMG -> BB]
  ground_truth:  [IMG -> BB]

  E2: AXIOM FORALL (d : IMG) : (in_training_set(d) = true)  => (cover (enlarge_by_post(dnn(d)), label(d)) = true)
  E3: AXIOM FORALL (d : IMG) : (in_training_set(d) = true) => (cover (label(d), ground_truth(d)) = true)
  E4: AXIOM FORALL (a : BB, b : BB, c : BB): ( cover(a, b) AND  cover (b, c) ) => (cover(a, c))

  E5: CONJECTURE FORALL (d: IMG) : (in_training_set(d) = true) => (cover(enlarge_by_post(dnn(d)), ground_truth(d)) = true)

  E1: AXIOM FORALL (behavior : BEHAVIOR, f1 : F1, f2: F2) : (FORALL (d : IMG) : (in_training_set(d) = true)  =>  behavior(f1(dnn(d)), f2(d)) = true) => (FORALL (d_op : IMG) : (in_odd(d_op) = true)  =>  behavior(f1(dnn(d_op)), f2(d_op)) = true)

  G1: CONJECTURE FORALL (d: IMG) : (in_odd(d) = true) => (cover(enlarge_by_post(dnn(d)), ground_truth(d)) = true)

END FASE22_Perception_full

\end{lstlisting}

\subsection{Part B: Proof in PVS}

Here we list how lemma~E5 is proved, followed by using E5 to prove~G1. 

\begin{lstlisting}
E5 :  
  |-------
{1}   FORALL (d: IMG):
        (in_training_set(d) = TRUE) =>
         (cover(enlarge_by_post(dnn(d)), ground_truth(d)) = TRUE)

Rule? (lemma E2)
Applying E2 
this simplifies to: 
E5 :  

{-1}  FORALL (d: IMG):
        (in_training_set(d) = TRUE) =>
         (cover(enlarge_by_post(dnn(d)), label(d)) = TRUE)
  |-------
[1]   FORALL (d: IMG):
        (in_training_set(d) = TRUE) =>
         (cover(enlarge_by_post(dnn(d)), ground_truth(d)) = TRUE)

Rule? (lemma E3)
Applying E3 
this simplifies to: 
E5 :  

{-1}  FORALL (d: IMG):
        (in_training_set(d) = TRUE) =>
         (cover(label(d), ground_truth(d)) = TRUE)
[-2]  FORALL (d: IMG):
        (in_training_set(d) = TRUE) =>
         (cover(enlarge_by_post(dnn(d)), label(d)) = TRUE)
  |-------
[1]   FORALL (d: IMG):
        (in_training_set(d) = TRUE) =>
         (cover(enlarge_by_post(dnn(d)), ground_truth(d)) = TRUE)

Rule? (lemma E4)
Applying E4 
this simplifies to: 
E5 :  

{-1}  FORALL (a: BB, b: BB, c: BB):
        (cover(a, b) AND cover(b, c)) => (cover(a, c))
[-2]  FORALL (d: IMG):
        (in_training_set(d) = TRUE) =>
         (cover(label(d), ground_truth(d)) = TRUE)
[-3]  FORALL (d: IMG):
        (in_training_set(d) = TRUE) =>
         (cover(enlarge_by_post(dnn(d)), label(d)) = TRUE)
  |-------
[1]   FORALL (d: IMG):
        (in_training_set(d) = TRUE) =>
         (cover(enlarge_by_post(dnn(d)), ground_truth(d)) = TRUE)

Rule? (skolem 1 ("d1"))
For the top quantifier in 1, we introduce Skolem constants: (d1),
this simplifies to: 
E5 :  

[-1]  FORALL (a: BB, b: BB, c: BB):
        (cover(a, b) AND cover(b, c)) => (cover(a, c))
[-2]  FORALL (d: IMG):
        (in_training_set(d) = TRUE) =>
         (cover(label(d), ground_truth(d)) = TRUE)
[-3]  FORALL (d: IMG):
        (in_training_set(d) = TRUE) =>
         (cover(enlarge_by_post(dnn(d)), label(d)) = TRUE)
  |-------
{1}   in_training_set(d1) =>
       cover(enlarge_by_post(dnn(d1)), ground_truth(d1))

Rule? (inst -2 "d1")
Instantiating the top quantifier in -2 with the terms: 
 d1,
this simplifies to: 
E5 :  

[-1]  FORALL (a: BB, b: BB, c: BB):
        (cover(a, b) AND cover(b, c)) => (cover(a, c))
{-2}  in_training_set(d1) => cover(label(d1), ground_truth(d1))
[-3]  FORALL (d: IMG):
        (in_training_set(d) = TRUE) =>
         (cover(enlarge_by_post(dnn(d)), label(d)) = TRUE)
  |-------
[1]   in_training_set(d1) =>
       cover(enlarge_by_post(dnn(d1)), ground_truth(d1))

Rule? (inst -3 "d1")
Instantiating the top quantifier in -3 with the terms: 
 d1,
this simplifies to: 
E5 :  

[-1]  FORALL (a: BB, b: BB, c: BB):
        (cover(a, b) AND cover(b, c)) => (cover(a, c))
[-2]  in_training_set(d1) => cover(label(d1), ground_truth(d1))
{-3}  in_training_set(d1) => cover(enlarge_by_post(dnn(d1)), label(d1))
  |-------
[1]   in_training_set(d1) =>
       cover(enlarge_by_post(dnn(d1)), ground_truth(d1))

Rule? (inst -1 "enlarge_by_post(dnn(d1))" "label(d1)" "ground_truth(d1)")
Instantiating the top quantifier in -1 with the terms: 
 enlarge_by_post(dnn(d1)), label(d1), ground_truth(d1),
this simplifies to: 
E5 :  

{-1}  (cover(enlarge_by_post(dnn(d1)), label(d1)) AND
        cover(label(d1), ground_truth(d1)))
       => (cover(enlarge_by_post(dnn(d1)), ground_truth(d1)))
[-2]  in_training_set(d1) => cover(label(d1), ground_truth(d1))
[-3]  in_training_set(d1) => cover(enlarge_by_post(dnn(d1)), label(d1))
  |-------
[1]   in_training_set(d1) =>
       cover(enlarge_by_post(dnn(d1)), ground_truth(d1))

Rule? (grind)
Trying repeated skolemization, instantiation, and if-lifting,
Q.E.D.


Installing rewrite rule sets.singleton_rew (all instances)
G1 :  

  |-------
{1}   FORALL (d: IMG):
        (in_odd(d) = TRUE) =>
         (cover(enlarge_by_post(dnn(d)), ground_truth(d)) = TRUE)

Rule? (lemma E1)
Applying E1 
this simplifies to: 
G1 :  

{-1}  FORALL (behavior: BEHAVIOR, f1: F1, f2: F2):
        (FORALL (d: IMG):
           (in_training_set(d) = TRUE) =>
            behavior(f1(dnn(d)), f2(d)) = TRUE)
         =>
         (FORALL (d_op: IMG):
            (in_odd(d_op) = TRUE) =>
             behavior(f1(dnn(d_op)), f2(d_op)) = TRUE)
  |-------
[1]   FORALL (d: IMG):
        (in_odd(d) = TRUE) =>
         (cover(enlarge_by_post(dnn(d)), ground_truth(d)) = TRUE)

Rule? (lemma E5)
Applying E5 
this simplifies to: 
G1 :  

{-1}  FORALL (d: IMG):
        (in_training_set(d) = TRUE) =>
         (cover(enlarge_by_post(dnn(d)), ground_truth(d)) = TRUE)
[-2]  FORALL (behavior: BEHAVIOR, f1: F1, f2: F2):
        (FORALL (d: IMG):
           (in_training_set(d) = TRUE) =>
            behavior(f1(dnn(d)), f2(d)) = TRUE)
         =>
         (FORALL (d_op: IMG):
            (in_odd(d_op) = TRUE) =>
             behavior(f1(dnn(d_op)), f2(d_op)) = TRUE)
  |-------
[1]   FORALL (d: IMG):
        (in_odd(d) = TRUE) =>
         (cover(enlarge_by_post(dnn(d)), ground_truth(d)) = TRUE)

Rule? (grind)
Trying repeated skolemization, instantiation, and if-lifting,
Q.E.D.


\end{lstlisting}

\end{document}